# Ultracompact Adiabatic Bi-sectional Tapered Coupler for the Si/III-V Heterogeneous Integration

**Qiangsheng Huang[1], Jianxin Cheng[2], Liu Liu[1, 2], and Sailing He[1, 2, 3,*]**

[1] *State Key Laboratory for Modern Optical Instrumentation, Centre for Optical and Electromagnetic Research, Zhejiang Provincial Key Laboratory for Sensing Technologies, Zhejiang University, East Building No.5, Zijingang Campus, Zhejiang University, Hangzhou 310058, China*

[2] *ZJU-SCNU Joint Research Center of Photonics, Centre for Optical and Electromagnetic Research, South China Academy of Advanced Optoelectronics, South China Normal University, 510006 Guangzhou, P. R. China*

[3] *Department of Electromagnetic Engineering, Royal Institute of Technology, 100 44 Stockholm, Sweden*

[*] *sailing@kth.se*

**Abstract:** An ultracompact bi-sectional adiabatic tapered coupler, which is suitable for an adiabatic mode transformation between a common single mode SOI wire waveguide and a Si/III-V hybrid waveguide, are proposed for Si/III-V heterogeneous integration. Since the bi-sectional tapered coupler mimics a semi-3D taper and avoids exciting the unwanted high-order modes in the thick p-InP cladding layer (which is removed in the first tapered section), the length of the adiabatic mode coupler can be dramatically shortened. Taking into account the mask design and the fabrication tolerance, we design the tapered structure only in the III/V structure, while keeping the SOI wire waveguide straight. In the proposed structure, the length of the bi-sectional tapered coupler can be 9.5 µm with a large fundamental mode-coupling ratio (over 95%) in a bandwidth of ~100 nm, and provides ±100 nm tolerance to misalignment, even when the BCB layer is as thick as 50 nm.

## References and links

## 1. Introduction

Photonics based on Si/III-V heterogeneous integration through the DVS-BCB adhesive wafer bonding method or molecular direct wafer bonding method [1,2] have been extensively investigated for many applications (e.g. inter-chip optical interconnects [1,2], free-space optical communication links [3]), due to its feasibility in combining active and passive photonic submodules onto one silicon substrate. Among various Si/III-V heterogeneously integrated devices, a common critical issue is to design a compact mode coupler structure to efficiently route light between the III-V active section (which is normally thick to achieve vertical electrical injection) and the silicon waveguide (which is normally thin for single-mode operation). Adiabatic taper has been considered to be an optimal structure for a mode converter due to its low insertion loss, low reflection losses [4], and good fabrication tolerance [5, 6]. However, in order to guarantee an adiabatic behavior, the tapered structure should be sufficiently long.

To overcome the drawback, optimizations of adiabatic tapered couplers have been developed in various ways. One way to reduce the adiabatic taper length is to use a multi-step or complex shape tapered structure in both the Si waveguide and the III-V waveguide [5, 6]. Another way is to use a multi-level tapered structure [4]. However, the taper lengths in Refs. [4-6] are still relatively long, i.e., over 20 μm. Ref, [7] further demonstrated the shortest possible length of an adiabatic mode coupler, under the assumption of only even and odd supermodes excited in the tapered structure. However, Ref. [7] did not give a specific tapered geometrical shape to achieve the shortest taper length. Recently, Refs. [8, 9] demonstrated compact tapered couplers by using a lateral-current-injection III-V structure or a III-V membrane, where a thick p-cladding does not exist. However, these kinds of tapered structures cannot be used in a traditional vertical-current-injection III-V structure.

In this letter, we propose an ultracompact bi-sectional adiabatic tapered coupler consisting of two linearly tapered sections only in the III-V structure, and the underneath SOI waveguide is kept straight with a common dimension of 600 nm × 220 nm. The bi-sectional tapered coupler mimics a semi-3D taper, which avoids exciting high-order modes in the thick p-cladding layer . In this way, the total taper length can be less than 10 μm even when the thickness of the bonding layer (defined as the distance between the top of the SOI structure and the bottom of the III-V structure) is relatively large, e.g., >50 nm. We also study the influence of the bonding layer thickness and misalignment on the performance of the proposed bi-sectional adiabatic tapered coupler.

## 2. Design of the bi-sectional tapered coupler

Fig. 1(a)-1(c) show a schematic structure of the proposed bi-sectional tapered coupler, as well as the cross-section of the Si/III-V hybrid waveguide. In the first taper section, the optical mode is converted from the silicon waveguide to the 0.598 μm thick III-V waveguide [including the first SCH (separate-confinement heterostructure) layer, MQW (multi-Quantum-Well) layer and the second SCH layer], without the thick p-cladding layer. At the end of the first taper, the optical mode is already well-matched with the final optical mode in the final Si/III-V hybrid waveguide. Thus, in the second taper section, the length can be very short. The thicknesses and the refractive indices of the materials composing the hybrid Si/III-V waveguide are shown in Fig. 1(a).

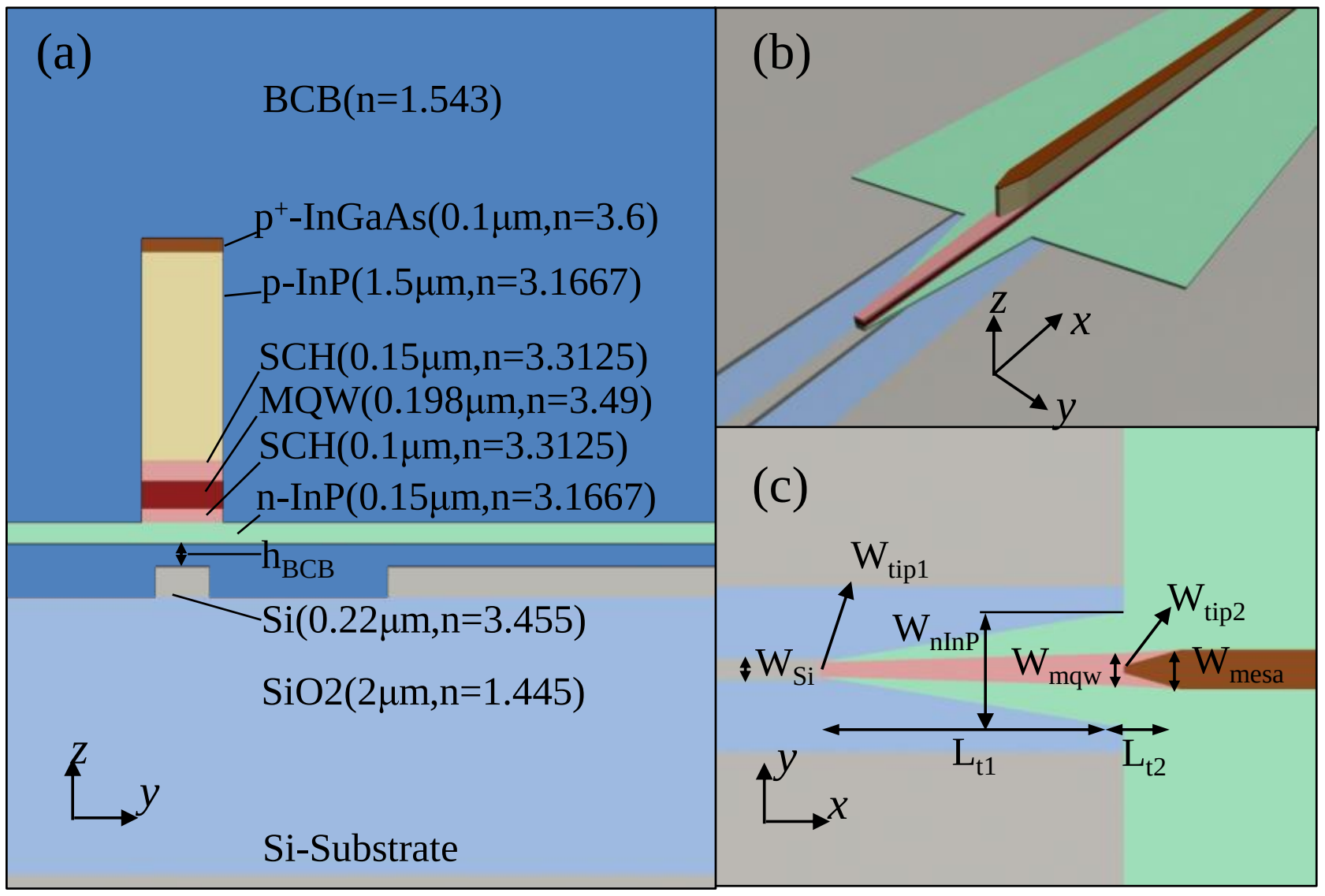

Fig.1. (a) Cross-sectional view of the hybrid Si/III-V waveguide. (b) 3D view and (c) top view of the bi-sectional adiabatic tapered coupler, respectively.

In order to reduce the complexity of the taper, we keep the silicon waveguide straight, without a tapered structure. This makes it more flexible in the post III-V fabrication process without considering the alignment of the tapered structure in the x direction (cf., Fig. 1). In the present design, the width ($W_{si}$) and the thickness of the SOI waveguide are chosen to be 600 nm and 220 nm respectively, which are values of a standard single-mode strip waveguide. We consider a bonding layer thickness ($h_{BCB}$) between 0 nm to 100 nm, which is a typical value for the adhesive or molecular bonding method. In this range, the final hybrid waveguide will present a confinement factor of 41%-44% in the MQW layer, which is appropriate for hybrid semiconductor lasers, SOAs, and modulators.

It is interesting to investigate how an adiabatic tapered coupler can be achieved to convert the mode between the silicon waveguide and hybrid waveguide, even the silicon waveguide is kept straight. The necessary condition to achieve an adiabatic mode transform is that there exists an intersection between the effective mode indices in the silicon waveguide and the III-V waveguide along the taper region [10]. Normally, the effective mode index of the III-V waveguide at the end of the taper is larger than the effective mode index of the silicon waveguide. Thus, the fundamental mode of the final hybrid waveguide will mainly reside in the III-V layer, which is a more desirable case. However, at the beginning of the taper, the fundamental mode should mainly remain in the silicon waveguide. This is achievable, as long as the width of the III-V waveguide at the beginning of the taper ($W_{tip1}$) is small enough. Thus, it is possible to design an adiabatic tapered coupler with a straight silicon waveguide.

In the present example, we choose the width of the active layer ($W_{mqw}$) at the end of the first taper to be 0.8 µm. The effective index of this III-V waveguide without the p-cladding layer is already well above the effective index of the silicon waveguide mode. This means that the adiabatic coupling between the silicon waveguide and the III-V waveguide mainly occurs in the first taper. At the end of the second taper, the width of the p-cladding layer and active layer ($W_{mesa}$) are expanded from $W_{tip2}$ to 1 µm, in order to enhance the optical confinement in the MQW layer.

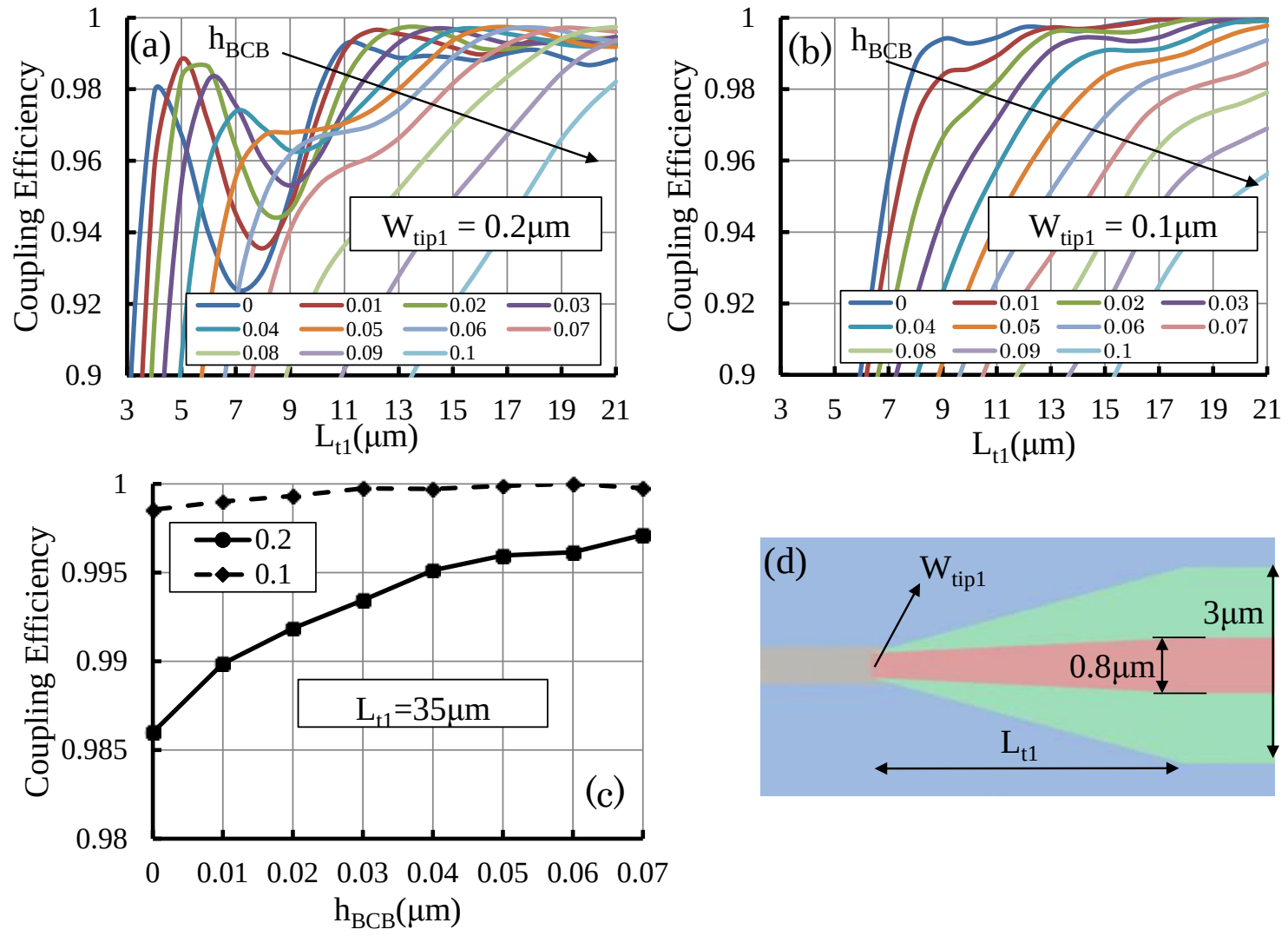

Fig. 2. The coupling efficiency as $L_{t1}$ varies under different thicknesses of BCB layer (from 0 to 0.1μm): (a) $W_{tip1}$ = 0.2 μm, (b) $W_{tip2}$ = 0.1 μm. (c) The stable coupling efficiency with $L_{t2}$ = 35 μm as $h_{BCB}$ varies. Solid and dashed lines in (c) show the results for $W_{tip1}$ = 0.2 μm and 0.1μm, respectively. (d) The structure parameter of the first section taper.

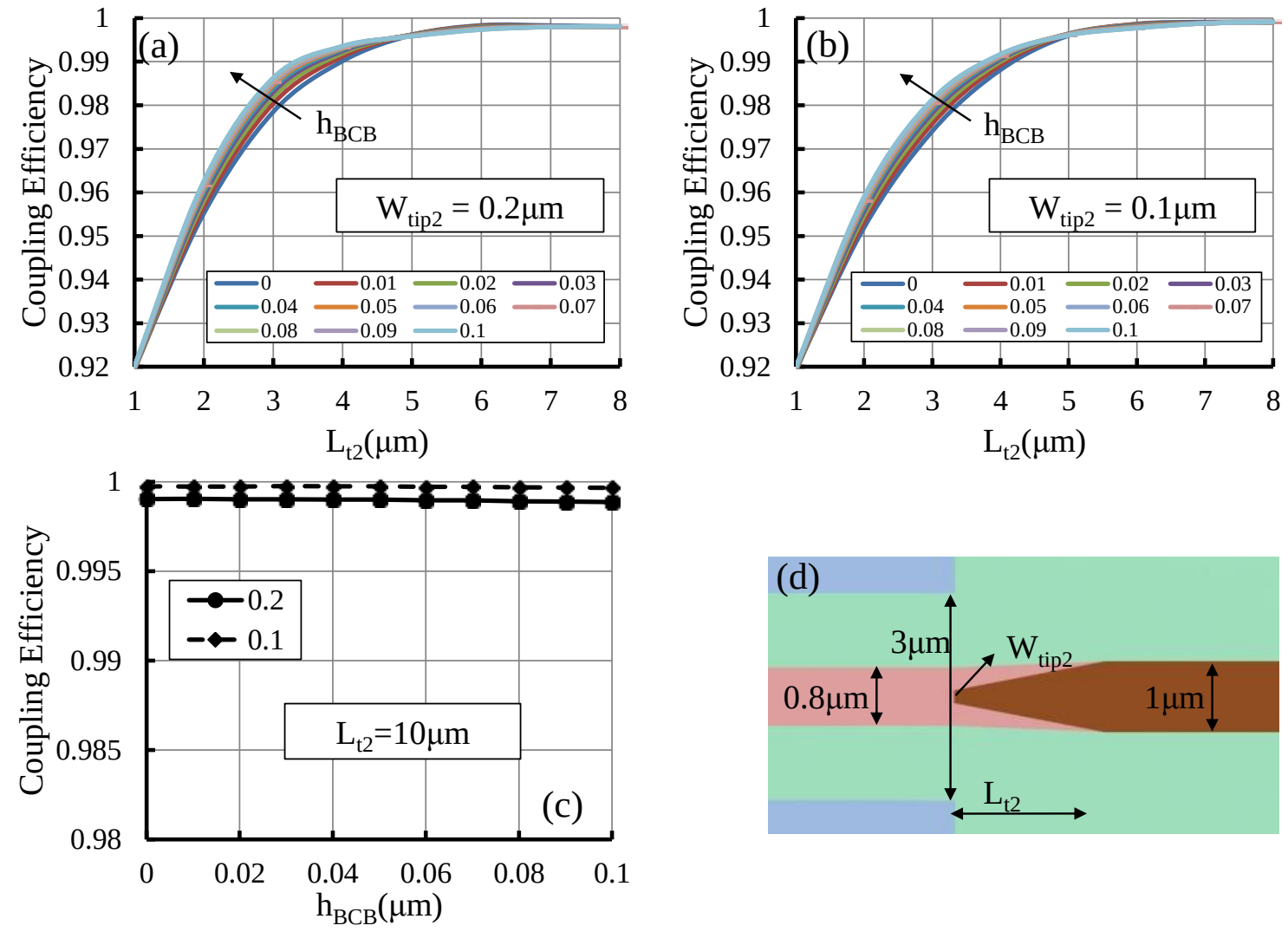

Fig.3. The coupling efficiency as $L_{t2}$ varies under different thicknesses of BCB layer (from 0 to 0.1μm): (a) $W_{tip2}$ = 0.2 μm, (b) $W_{tip2}$ = 0.1 μm. (c) The stable coupling efficiency with $L_{t2}$ = 10 μm as $h_{BCB}$ varies, where the solid and dashed lines in (c) show the results for $W_{tip2}$ = 0.2 μm and 0.1μm, respectively. (d) The structure parameter of the second section taper.

We first analyze the performance of the first taper with different values for $L_{t1}$, $h_{BCB}$, and $W_{tip1}$, using a three-dimensional finite-difference time-domain method [11]. Fig. 2(a) and 2(b) show the coupling efficiency as a function of the length of the first taper length ($L_{t1}$), when $h_{BCB}$ and $W_{tip1}$ vary. The coupling efficiency is defined as the fraction of power coupled to the fundamental mode of the output waveguide at the end of the taper. Note the width of the n-InP layer (in green color in Fig. 2(d)) keeps constant after the first taper in order to simplify the analysis (the reflection due to the geometrical mismatch at the end of the first taper is included in Fig. 3 below). The coupling loss mainly has two causes. The first is the coupling to the higher order mode at the entrance of the taper, since the taper tip cannot be infinitely small due to fabrication issues. The second comes from the insufficient taper length. In this case, the coupling to the higher order mode can also occur when light travels along the taper. When $W_{tip1}$ = 0.2 µm, and $h_{BCB}$ < 0.04 µm, the coupling efficiency has a large fluctuation with $L_{t1}$ < 11 µm, and this would require the adiabatic tapered length to increase. This fluctuation is mainly caused by the coupling between the even and odd (the first higher-order) supermodes. Here, the even supermode is the fundamental mode which should adiabatically follow the taper. For a wider taper tip or a thinner bonding layer, the odd supermode gets excited more easily at the entrance of this taper. When the taper is not long enough, i.e., non-adiabatic, these two modes can couple between each other along the taper, which induces a large fluctuation, as shown in Fig. 1(a). Care must be taken when designing a taper working in this regime, since the performance may be sensitive to the structural parameters, and thus vulnerable to fabrication uncertainties. This fluctuation is eliminated as $h_{BCB}$ increases or $W_{tip1}$ decreases. However, a thicker BCB bonding layer also decreases the coupling strength between the SOI and III-V waveguide mode, and increases the taper length. Fig. 2(c) shows the stable coupling efficiency when the taper length is long enough (i.e. 35µm), to ensure that the taper is adiabatic for $h_{BCB}$ below 70 nm. It is shown that when $W_{tip1}$ = 0.2 µm, the stable coupling efficiency for $h_{BCB}$=0 nm is 98.6% and it increases as $h_{BCB}$ increases. In contrast, when $W_{tip1}$ = 0.1 µm, the stable coupling efficiency is nearly 100% even for $h_{BCB}$ = 0 nm. This is an indication of, as mentioned above, the larger coupling to the higher–order mode at the entrance of the taper for a wider $W_{tip1}$ or thinner $h_{BCB}$.

Fig. 3(a) and 3(b) show the coupling efficiency in the second taper with $W_{tip2}$ = 0.2 µm and 0.1 µm, respectively (the widths of the active waveguide and the n-InP layer at the entrance are fixed to 0.8 µm and 3 µm, respectively). From these figures, one can see that the coupling efficiency is insensitive to both the thickness of the BCB layer and $W_{tip2}$. The taper length required to achieve a high coupling efficiency is also very short, as the mode input to this taper is already well-matched to that in the final Si/III-V hybrid waveguide. Moreover, Fig. 3(c) shows that the stable coupling efficiency reaches over 99% with only a 10 µm long taper for different values of $h_{BCB}$ and $W_{tip2.}$

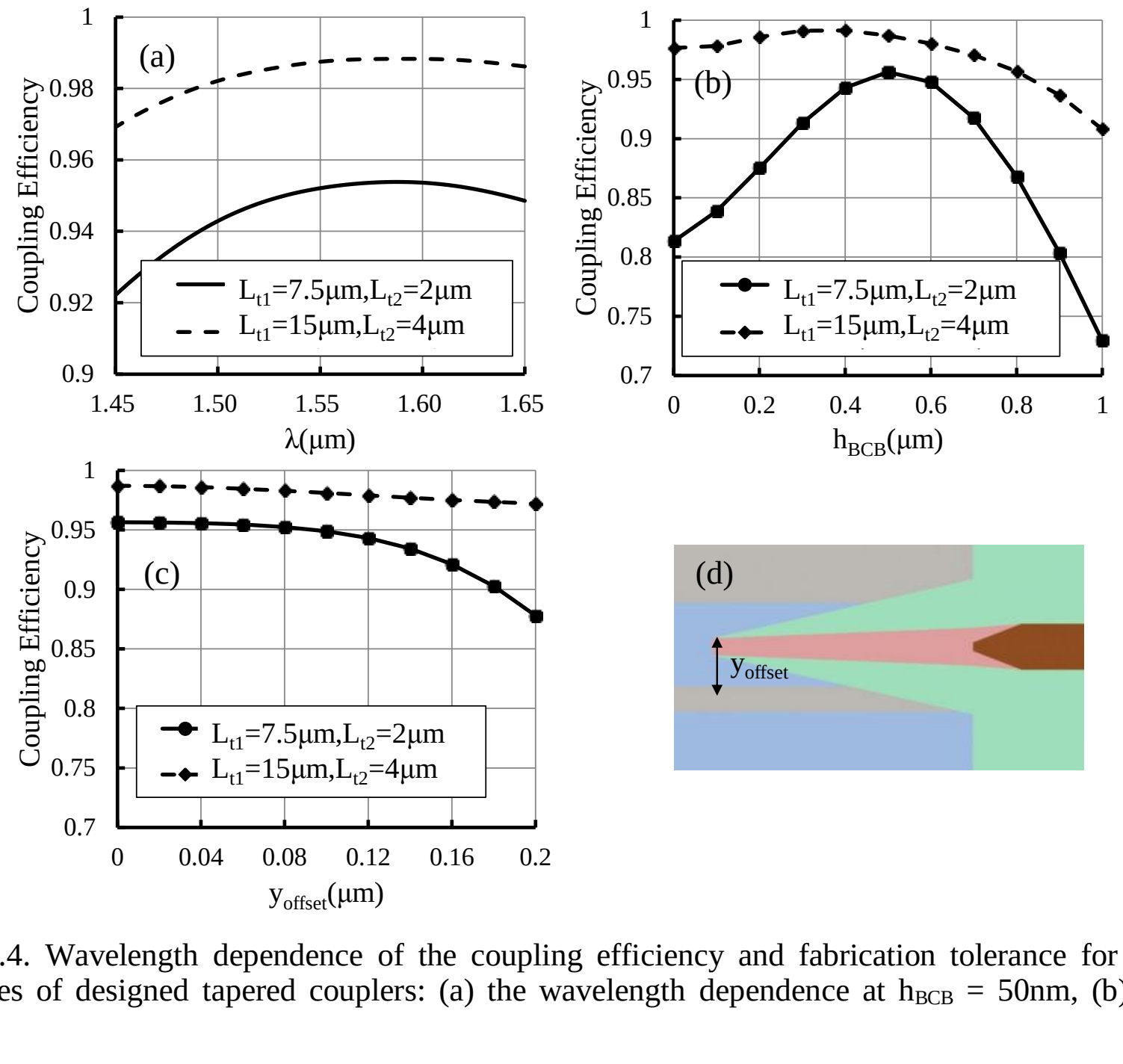

Fig.4. Wavelength dependence of the coupling efficiency and fabrication tolerance for two types of designed tapered couplers: (a) the wavelength dependence at $h_{BCB}$ = 50nm, (b) the variation of the thickness of the BCB layer at $\lambda$ = 1.55μm, (c) coupling efficiency in the presence of misalignment between the silicon waveguide and III-V waveguide ($y_{offset}$) with $h_{BCB}$ = 50nm and $\lambda$ = 1.55μm. (d) The structure parameter of a misaligned case.

In the present paper, we propose two bi-sectional tapered couplers, with both $W_{tip1}$ and $W_{tip2}$ equal to 0.2 μm. For the first tapered coupler, we choose $L_{t1}$ = 15 μm and $L_{t2}$ = 4 μm, and thus the total taper length is 19 μm. From Fig 4(a), one can see that this design gives a high coupling efficiency of over 98% at $h_{BCB}$ = 50 nm for the wavelength range from 1500 nm to 1650 nm. Fig 4(b) and (c) indicate that this 19 μm long taper also has a large fabrication tolerance to the variation of $h_{BCB}$ and the misalignment between the silicon waveguide and III-V waveguide. When $h_{BCB}$ varies from 0 nm to 80 nm, the coupling efficiency is still over 95% at $\lambda$ = 1.55 μm, meaning that this 19 μm long taper can be applied to a relatively thick bonding layer.

On the other hand, another ultracompact tapered coupler is also proposed with $L_{t1}$ = 7.5 μm and $L_{t2}$ = 2.5 μm. In this case, the total taper length is only 9.5 μm, and the coupling efficiency of the fundamental mode can still be over 95% when $h_{BCB}$ = 50 nm. The reflection of this tapered coupler from fundamental mode to (the counter-propagating) fundamental mode is lower than -40 dB when $h_{BCB}$ varies from 0 nm to 100 nm. The bandwidth and the fabrication tolerance of the designed 9.5-μm-long tapered coupler are also analyzed, as shown in Fig. 4(a)-4(c). This ultracompact coupler demonstrates a coupling efficiency of over 95% in a wavelength range of ~100 nm, and a misalignment tolerance of 100 nm. Although it is more sensitive to the variation of $h_{BCB}$, the coupling efficiency is still above 80% when $h_{BCB}$ varies between 0 nm to 80 nm. This ultracompact coupler is more suitable for a hybrid modulator [12], where the required coupling efficiency is not as high as a hybrid laser or SOA.

The light propagation from the silicon waveguide to the III-V waveguide for this ultracompact taper, with $h_{BCB}$ = 50 nm, is shown in Fig. 5(a). The mode profiles at the end of the first and the second taper section are shown in Fig. 5(b) and 5(c), respectively. One can see that the power coupling from silicon waveguide to III-V waveguide mainly occurs in the first taper section. As a comparison, in Fig. 6(a)-6(c), a conventional single-section taper design with the same taper parameters is also simulated. The only difference is that the thick p-cladding layer covers the whole taper. The coupling efficiency in this case is only 83%. One can clearly see that high-order modes are excited in the thick p-InP cladding layer along this tapered coupler, and thus degenerates the performance of the coupler.

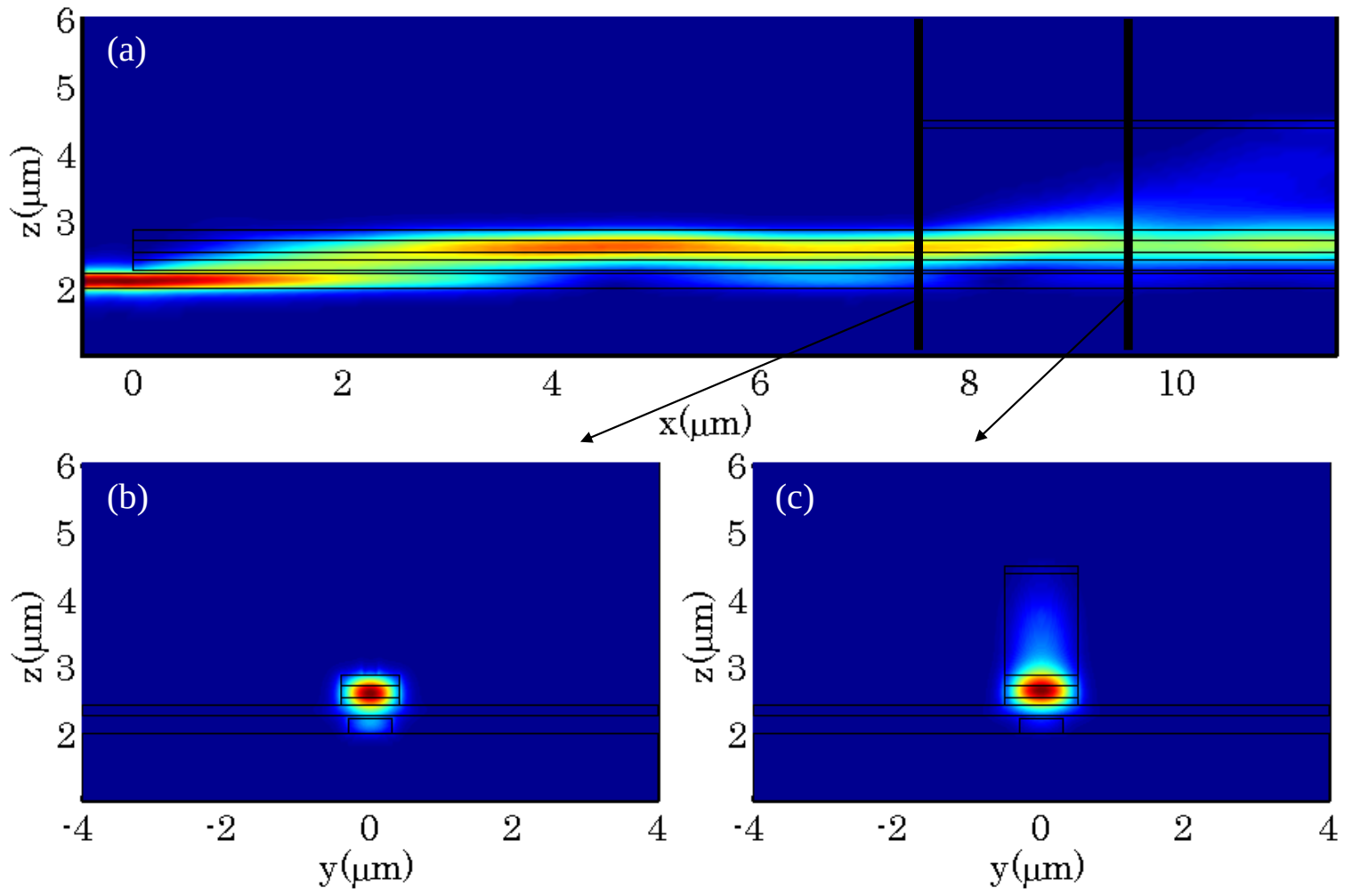

Fig.5. (a) Mode transformation in the designed bi-sectional tapered coupler with $L_{t1}$ = 7.5 μm, $L_{t2}$ = 2 μm, and $h_{BCB}$ = 50 nm. (b) Field profile at the end of the first tapered coupler. (c) Field profile at the end of the second tapered coupler.

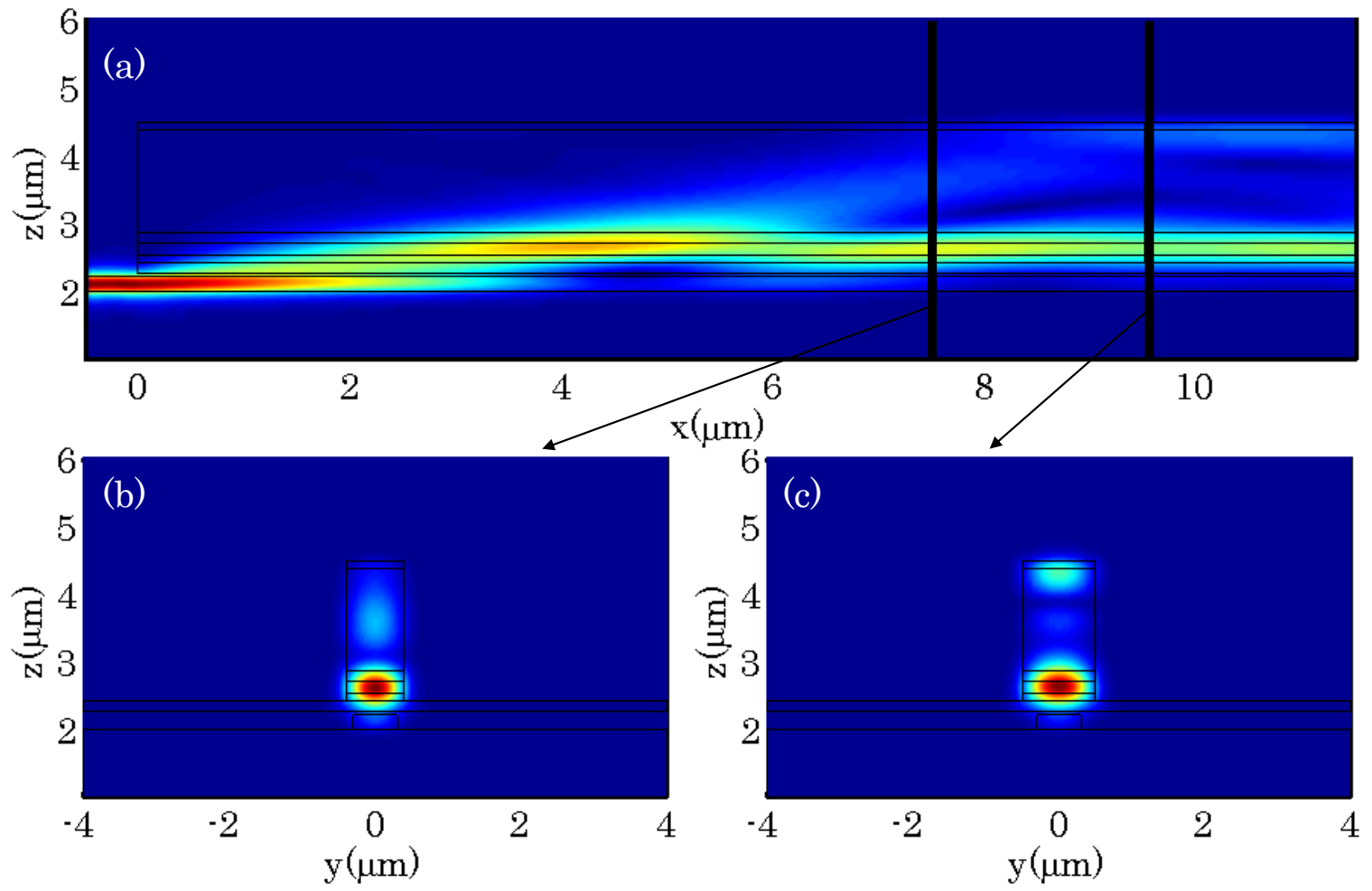

Fig.6. Comparison with a conventional tapered coupler with the p-cladding throughout the whole tapered structure. (a) Mode transformation. (b) Field profile at the end of the first tapered coupler. (c) Field profile at the end of the second tapered coupler.

## 3. Conclusions

In conclusion, an ultracompact bi-sectional tapered coupler with a standard straight SOI strip waveguide underneath has been proposed for adiabatic mode transformation between a common single mode SOI wire waveguide and a Si/III-V hybrid waveguide in either molecular bonding or DVS-BCB adhesive bonding technology. In the first tapered section, the thick p-cladding layer is removed, which avoids exciting high-order modes. In this way, a coupling efficiency of over 95% can be achieved in a wavelength range of ~100 nm even when the total taper length is only 9.5 μm. Our simulation has also shown that this 9.5 μm long taper provides ±100 nm tolerance to misalignment. We believe this ultracompact bi-sectional tapered coupler can be used in various Si/III-V heterogeneous integrated devices.

## Acknowledgment

This work is partially supported by the National High Technology Research and Development Program (863) of China (2012AA012201), National Nature Science Foundation of China (#61107020). We thank for valuable discussions from Yaocheng Shi, Keqi Ma and Yingchen Wu.